\documentstyle[12pt,epsf]{article}
\begin{document}
\def\be{\begin{equation}}
\def\ee{\end{equation}}
\def\bea{\begin{eqnarray}}
\def\eea{\end{eqnarray}}
\def\ti{\tau_{int}}
\def\mul{{\cal O}(e^{-\mu L})}
\title{
\bf \mbox{}  \\ A test of a new simulation algorithm for dynamical quarks.}
\vspace{.5cm}
\author{
{\bf T.D.Bakeyev}
\vspace{.5cm}
\date{}
\\ {\it Moscow State University}
}
\maketitle


\begin{abstract}

Some results of test runs on a $6^3\times 12$ lattice
with Wilson quarks and gauge group SU(2) for a previously
proposed fermion algorithm by A. Slavnov are presented.

\end{abstract}
\input epsf

\section{Introduction.}

The incorporation of fermionic degrees of freedom in lattice simulations
presents a serious difficulties which stem from their being anticommuting
variables. For the most actions in use which are quadratic
in fermionic fields, one can eliminate fermions by an analytic integration.
Unfortunately, the resulting expressions involve the determinant
of a very large matrix, which induces nonlocal long-range interactions
on the gauge fields, making the numerical simulations rather expensive.

The most often used method to simulate fermionic field theories
is now Hybrid Monte Carlo (HMC) algorithm \cite{HMC}, which addresses
the problem of nonlocality by linearizing the action in a succession
of molecular dynamics steps. It is exact and easily implementable,
but still the fermionic simulations turn out to be very time consuming
and alternative approaches are welcome.

An interesting idea has been put forward by M. L\"{u}scher \cite{Luescher},
who proposed
to approximate the inverse of the fermion matrix with a polynomial,
and then interprete the determinant as the partition function
of a local bosonic system. This technique has been studied thorougly
for the last five years and it was claimed to be competitive
to the HMC algorithm
and even superior when fermions are heavy
(a detailed analysis can be found in Refs\cite{K4,Forcrand,Forc4}).
However the efficiency of L\"{u}escher's method is abated
when quark mass decreases.

Another approach to the problem of dynamical fermions simulation
was proposed by A. Slavnov in papers \cite{Slalg1,Slalg2}, where
a D-dimensional fermion determinant was presented as a path integral of
a (D+1)-dimensional local bosonic action. In Ref.\cite{Toy1d} this
procedure was tested by numerical simulation of a one dimensional toy model
and it was shown that correct and accurate results can be obtained with a
reasonable size of lattice in auxiliary dimension. In this paper we
present some results from test runs for SU(2) QCD with 2 degenerate
heavy Wilson quarks,
using the version of Slavnov's algorithm from Ref.\cite{Slalg2}.

\section{The local bosonic algorithm.}

For the local gauge theory with two flavours of Wilson quarks
the effective distribution of the gauge field $U$ is given by

\be P[U]\propto det (D[U]+m)^2 e^{-S_g[U]}\label{1}\ee
where $D[U]$ is Wilson-Dirac operator

\be D=\frac{1}{2}\sum_\mu[\gamma_\mu(\nabla_\mu+\nabla_\mu^*)
-\nabla_\mu^*\nabla_\mu]\ee
$S_g[U]$ denotes the usual plaquette action and $m$ is the bare quark mass
which is related to hopping parameter $k$ through $k=1/(8+2m)$.
The lattice spacing has been set equal to unity for simplicity.
It will be advantageous for us to work with hermitean operator
$B=\gamma_5(D+m)$, so we rewrite expression (\ref{1}) in the form

\be P[U]\propto det B[U]^2 e^{-S_g[U]}\label{3}\ee

Now following Ref.\cite{Slalg2}, we introduce five dimensional
bosonic fields $\phi(x,t)\equiv\phi_n(x)$, where the extra coordinate
$t$ is defined on one dimensional chain of the length $L$ with the lattice
spacing $b$

\be t=nb\;\; ;\;\;  0\le n\le N\;\; ;\;\; L=Nb \ee
and four dimensional bosonic fields $\chi(x)$. After that we present
the determinant in eq.(\ref{3}) in the following form

\be det B[U]^2= \lim_{\mu\rightarrow 0}det(B[U]^2+\mu^2)
=\lim_{\mu\rightarrow 0}\;\;\lim_{b\rightarrow 0 ;L\rightarrow\infty}
\int e^{-S_{\mu,b,L}[U,\phi,\chi]}D\phi^* D\phi D\chi D\chi\label{5}\ee
where $S_{\mu,b,L}$ is a local bosonic action

\bea &&S_{\mu,b,L}[U,\phi,\chi]=\sum_x \sum_{n=1}^{N-1}\Bigl[
-(\phi_{n+1}^*(x)\phi_n(x)+h.c.-2\phi_n^*(x)\phi_n(x))
+\nonumber\\&&
+b(\imath\phi_{n+1}^*(x)B\phi_n(x)+h.c)
+b^2\phi_{n}^*(x)B^2\phi_n(x)
+\nonumber\\&&
+b\exp\{-\mu bn\}(\chi^*(x)(\mu+\imath B)\phi_n(x)+h.c.)\Bigr]
+\frac{1}{2\mu b}\sum_x\chi^*(x)\chi(x)\label{6}\eea
and the free boundary conditions in $t$ for the fields $\phi_n$ are imposed

\be \phi_0=0 \;\; ;\;\; \phi_N=0\ee

Let us prove eq.(\ref{5}), which shows that lattice QCD is a limit of purely
bosonic theory (see Ref.\cite{Slalg2}). The bosonic action (\ref{6}) is
a linearized version of the expression in the exponent of the
following integral

\bea && I_{\mu,b,L}[U]=\int \exp\Bigl\{\sum_\alpha \sum_{n=1}^{N-1}\Bigl[
(\phi_{n+1}^{\alpha *}\exp\{-\imath B_\alpha b\}\phi_n^\alpha
+h.c.-2\phi_n^{\alpha
*}\phi_n^\alpha)-\nonumber\\&&-b\exp\{-\mu bn\} (\chi^{\alpha
*}(\mu+\imath B_\alpha)\phi_n^\alpha+h.c)\Bigr]-\frac{1}{2\mu b}
\sum_\alpha\chi^{\alpha *}\chi^\alpha\Bigr\}
D\phi^* D\phi D\chi^* D\chi\label{8} \eea
where instead of $x$-representation we used a basis formed by the eigenvectors
of the operator $B$, $B_\alpha$ being corresponding eigenvalues.
Indeed, expanding $\exp\{-\imath B_\alpha b\}$ in eq.(\ref{8})
in a Taylor series, keeping only the terms of the order ${\cal O}(b^2)$
and substituting
\be \frac{b^2}{2}(\phi_{n+1}^{\alpha *}B_\alpha^2\phi_n^\alpha+h.c.)
\rightarrow b^2\phi_n^{\alpha *}B^2_\alpha\phi_n^\alpha \label{8a}\ee
we get the expression (\ref{6}). The substitution (\ref{8a}) also introduces
corrections of the order ${\cal O}(b^2)$ and ensures that
the action (\ref{6}) is strictly positive.

By changing variables
\be \phi_n^\alpha\rightarrow\exp\Bigl\{-\imath B_\alpha nb\Bigr\}
\phi_n^\alpha \ \ ;\ \ \phi_n^{\alpha *}\rightarrow\exp\Bigl\{\imath
B_\alpha nb\Bigl\}\phi_n^\alpha \label{9}\ee
we can rewrite the equation (\ref{8}) as the gaussian integral over
$\phi$ with the quadratic form which does not depend on $B_\alpha$

\bea && I_{\mu,b,L}[U]=\int \exp\Bigl\{\sum_\alpha \sum_{n=1}^{N-1}\Bigl[
(\phi_{n+1}^{\alpha *}\phi_n^\alpha+h.c.-2\phi_n^{\alpha
*}\phi_n^\alpha)-\nonumber\\&&-b(\exp\{-(\mu+\imath B_\alpha) bn\}
\chi^{\alpha*}(\mu+\imath B_\alpha)\phi_n^\alpha+h.c)\Bigr]-\frac{1}{2\mu b}
\sum_\alpha\chi^{\alpha *}\chi^\alpha\Bigr\}
D\phi^* D\phi D\chi^* D\chi\nonumber\\&&\label{10} \eea

Let us integrate the expression (\ref{10}) over the fields $\phi$.
To do so it is sufficient to find a stationary point of the exponent.
For small $b$ the sum over $n$ can be replaced by the integral over
continuous variable $t=nb$ and the equations for the stationary point acquire
the form

\bea && \ddot \phi^\alpha -b^{-1}\chi^\alpha (\mu-\imath B_\alpha)
e^{-(\mu-\imath B_\alpha)t}=0 \nonumber\\&&
\phi^\alpha(0)=\phi^\alpha(L)=0\eea
The replacement of the sum by the integral also introduces corrections of order
$O(b^2)$.
The solutions of these equations are:
\be \phi^\alpha (t)=\frac{\chi^\alpha}{b(\mu-\imath B_\alpha)}
\Bigl(e^{-(\mu-\imath B_\alpha)t}+\frac{t}{L}(1-e^{-(\mu-\imath
B_\alpha )L})-1\Bigr)\ee

Substituting these solutions to the integrand (\ref{10}),
integrating over $t$ and rescaling the fields
$\;\chi\rightarrow \sqrt{bL}\chi\;\;$,
we get

\be \lim_{b\rightarrow 0}I_{\mu,b,L}[U]=\int\exp\Bigl\{-\sum_\alpha \frac{
\chi^{\alpha *}\chi^\alpha}{\mu^2+B_\alpha^2}(1-2e^{-\mu L}\cos{B_\alpha L}
+e^{-2\mu L})\Bigr\}D\chi^* D\chi \label{13}\ee
Therefore:
\be \lim_{b\rightarrow 0,L\rightarrow\infty}I_{\mu,b,L}[U]=det(B[U]^2+\mu^2)
\label{14}\ee

The equality (\ref{5}) is proven.
For finite $b,L$ and $\mu$ this equation is approximate

\be DetB[U]^2\approx\int e^{-S_{\mu,b,L}[U,\phi,\chi]}
D\phi^* D\phi D\chi^* D\chi\label{15}\ee
and has to be corrected by the terms

\be O(b^2)\;\; ;\;\; O(e^{-\mu L})\;\; ;\;\; {\cal O}(\mu^2)
\label{16}\ee

It was shown that lattice QCD is a limit of a purely bosonic local
theory with the action

\be S_{eff}[U,\phi,\chi]=S_g[U]+S_{\mu,b,L}[U,\phi,\chi]\label{17}\ee
Making use of this action, we can simulate the theory by locally updating
the boson fields $U,\phi$ and $\chi$.

Choosing the number of points in auxiliary dimension $N$ we
can fully control the systematic errors of the algorithm (\ref{16}).
Indeed, fixing $\mu L=\epsilon$ we have

\be N=\frac{\epsilon}{b \mu}\ee
Increasing $N$ we can increase $\epsilon$ and decrease $b$ and $\mu$,
making the errors (\ref{16}) small enough in comparison with an available
level of statistical precision. Since
the computational cost of the algorithm depends on $N$,
the parameters $\epsilon,\mu,b$ should be optimally tuned to make $N$
as small as possible.

\section{Test results. Investigation of systematic errors of the
algorithm.}

In this section we present the results from numerical simulations
of the bosonic
theory (\ref{17}) on a $6^3\times 12$ lattice with gauge group SU(2)
and bare parameters $\beta=2.12$ and $k=0.15$.
The simulations for the same model and parameters were already done
in Ref.\cite{LuescherE} within the framework of Luescher's formulation
of the dynamical quarks problem \cite{Luescher}. That will allow
us to compare the algorithm explored in this paper
with the Luescher's one.


The implementation was done on a Quadrics Q4 machine (APE computer)
with 32 nodes. A full update cycle involved 1 heatbath and 1 overrelaxation
sweep for the fields $\phi_n$ followed by 1 heatbath sweep for the fields
$\chi$ and gauge fields $U$. (Our preparatory tests showed that overrelaxing
the gauge fields and the fields $\chi$ does not decrease autocorrelaton times
$\tau_{int}$. On the contrary, overrelaxing the fields $\phi_n$
can decrease $\tau_{int}$ substantially.) For each set of parameters
$\mu,b,N$ we have performed at least 30000 and sometimes up to 80000
such cycles after thermalization of the system, using the random number
generator of Ref.\cite{randl}.

Our test runs were structurally divided into a sequence of "experiments"
in which the systematic errors (\ref{16}) were studied separately.
In each "experiment" we fixed some of the parameters
$\mu,b,\mu L$, holding the corresponding errors fixed and
altered the other parameters. This way of investigation of the systematic
errors allows to determine the values of parameters at which
one may be able to get correct results at given level of statistics,
using the bosonic theory with relatively small number of points
in auxiliary dimension $N$.

In the first "experiment" we fixed $\mu L=4.0$, $\mu=1.0$ to study
the alteration of the plaquette value when the lattice spacing
in auxiliary dimension $b$ changes. By doing so we investigated
the algorithm's
error of the order ${\cal O}(b^2)$ for fixed errors $\mul$ and
${\cal O}(\mu^2)$. The results of the simulations
are reported in Table 1. In Fig. 1 the plaquette values multiplied by 100
are plotted against the inverse lattice spacing $b^{-1}$.
The plot shows that the plaquette value stabilizes quickly when $b$
decreases. From the data obtained we see that at this level of statistics
one may be able to do with $b=0.08$ to make the systematic error of the order
${\cal O}(b^2)$ small.

\vspace{1cm} \begin{center}
\begin{tabular}{|l|l|l|l|l|l|l|} \hline
N         & 20& 30& 40& 50 & 60& HMC\\
\hline
$b^{-1}$  & 5 &7.5& 10&12.5& 15& -  \\
\hline
plaquette &0.5467(2)&0.5697(3)&0.5782(4)&0.5807(8)&0.5812(12)&0.5796(6)\\
\hline
$\ti $&24(2)&67(8)&94(18)&*&*& - \\
\hline
\end{tabular}\label{table1}
\end{center}
{\bf Table 1:} {\small Simulation results for the plaquette at
$\mu=1.0$ and $\mu bN=4.0$. The autocorrelation time for the plaquette
is measured in unit of iteration sweeps. A star means that the autocorrelation
time was too long to be measured accurately. In the last column
we adduce for comparison a value obtained from the Hybrid Monte Carlo
for the same lattice and parameters $\beta$, $k$ in Ref.\cite{LuescherE}.}
\vspace{1cm}

\begin{figure}
\epsfxsize=0.47\textwidth
\epsfbox{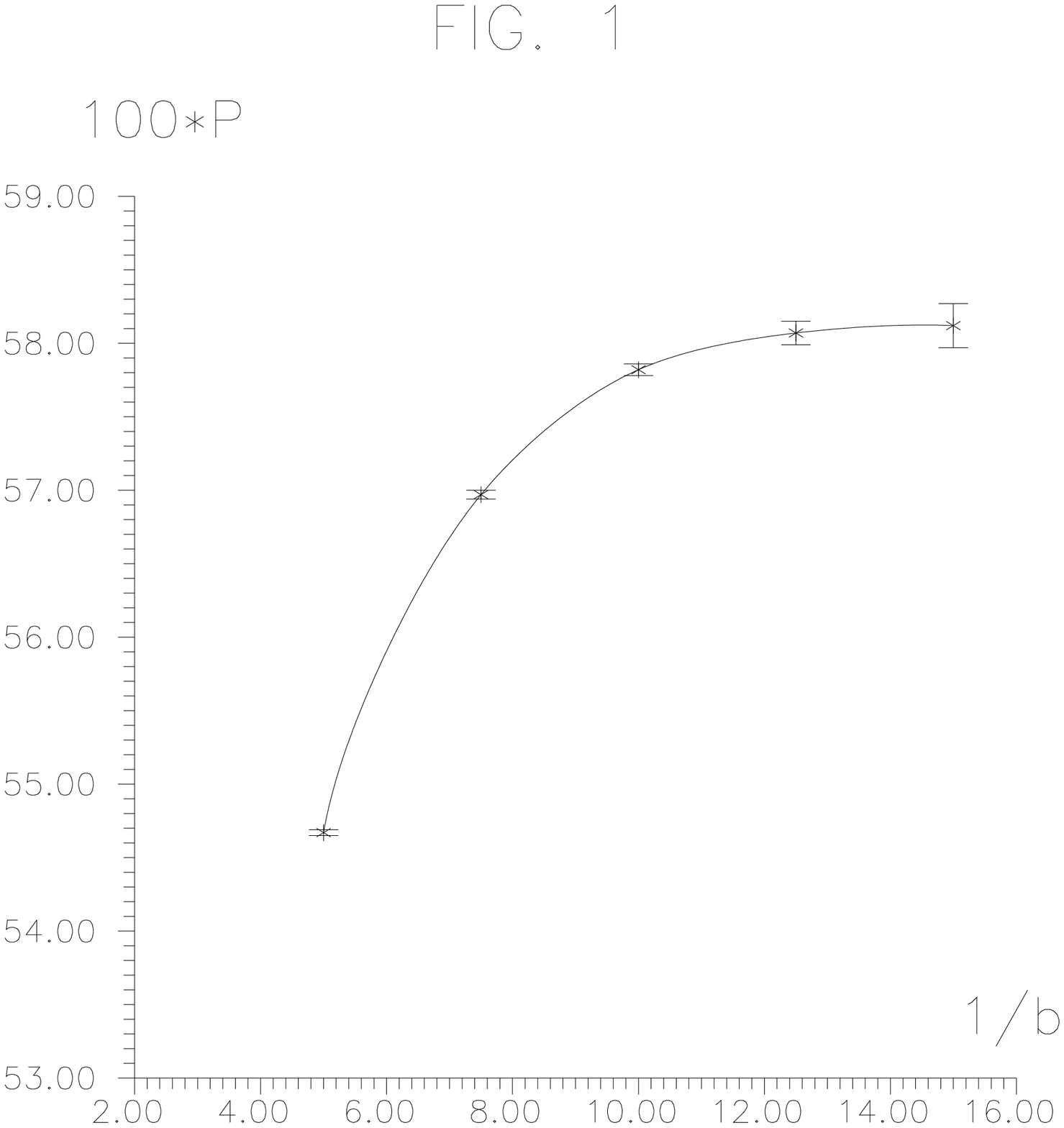}
\end{figure}

In the second "experiment" we attempted to investigate
the character of systematic deviations of the order ${\cal O}(e^{-\mu L})$.
For that purpose we fixed $\mu=1$ and $b=0.1$
in order to fix the deviations connected with the errors of the order
${\cal O}(\mu^2)$ and ${\cal O}(b^2)$.
As a work hypothesis we assumed that the correction terms of the order
${\cal O}(b^2)$ to the approximation
(\ref{15}) do not depend on $N$ at fixed $b$ (it will be seen from the
simulation results that this hypothesis is wrong).

The results are reported in Table 2. In Fig. 2 the plaquette values multiplied
by 100 are plotted against the number of points in auxiliary dimension $N$.

\vspace{1cm} \begin{center}
\begin{tabular}{|l|l|l|l|l|} \hline
N& 20& 30& 40& 50 \\
\hline
$\mu L$& 2& 3& 4& 5 \\
\hline
plaquette&0.5833(1) &0.5790(3) &0.5782(4) &0.5755(8)\\
\hline
$\ti$&28(2)&65(8)&94(18)& * \\
\hline
\end{tabular}\label{table2}
\end{center}
{\bf Table 2:} {\small Simulation results at $\mu=1.0$ , $b=0.1$.}
\vspace{1cm}

\begin{figure}
\epsfxsize=0.5\textwidth
\epsfbox{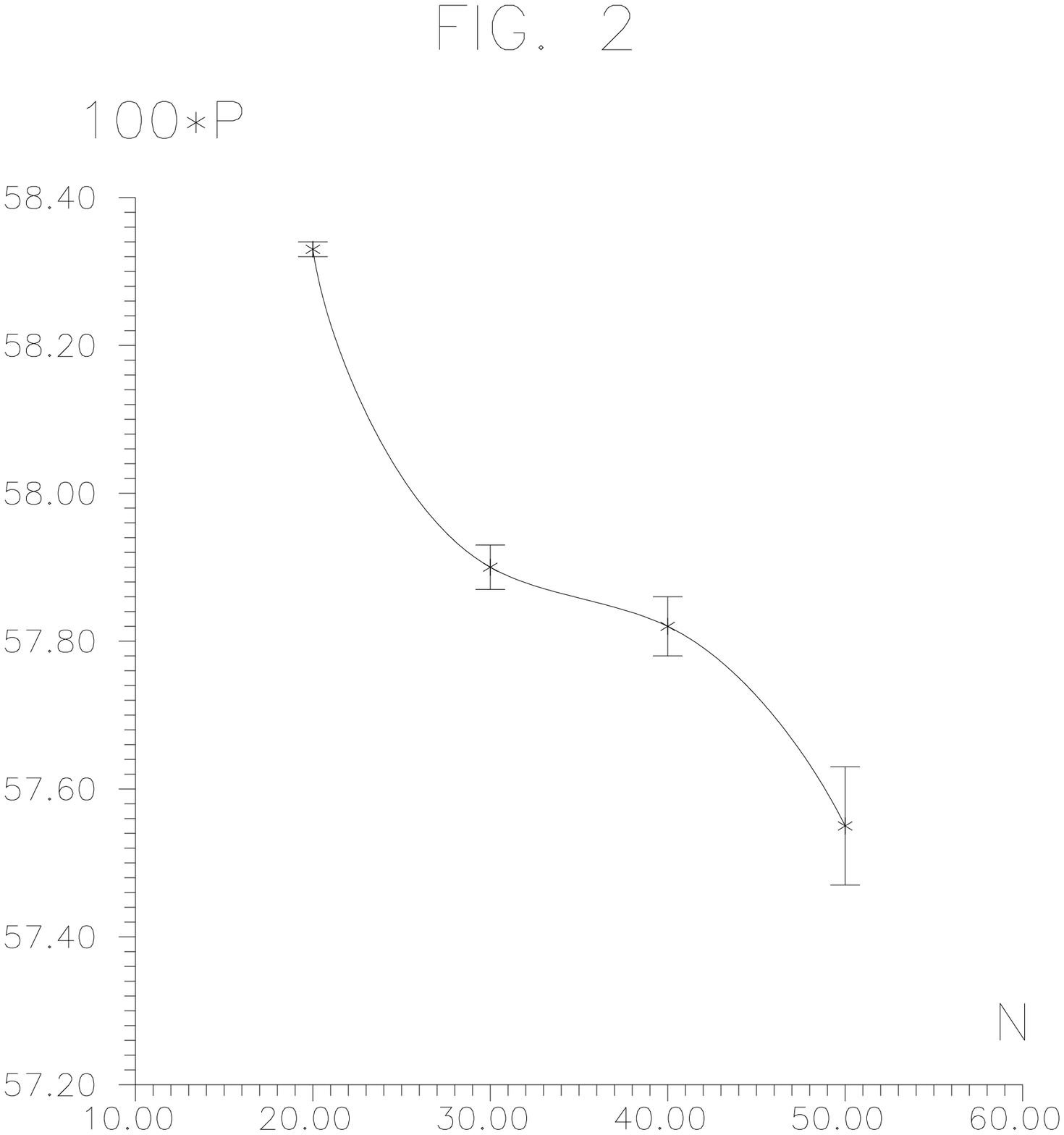}
\end{figure}

The plot shows that the plaquette value does not stabilize when $N$ increases.
Since the error of the order ${\cal O}(\mu^2)$ can not depend on
$N$ (the approximation $DetB^2\approx Det (B^2+\mu^2)$ was used even
before the bosonization procedure and introduction of auxiliary dimension),
the only explanation to this result is the dependence on $N$ of the error
${\cal O}(b^2)$ for fixed $b$.

To demonstrate this on a theoretical level, let us return to the expression
(\ref{10}) and integrate it over $D\phi_n$ for finite $b$ (i.e. not using
the simplifying assumption $b\rightarrow 0$).
Then the equations for the stationary point
acquire the form

\bea && 2\phi^\alpha_n-\phi^\alpha_{n+1}-\phi^\alpha_{n-1}+
b\chi^\alpha (\mu-\imath B_\alpha)e^{-(\mu-\imath B_\alpha)nb}=0
\nonumber\\&&
\phi_0=\phi_N=0\eea

The solutions of these equations are
\be \phi^\alpha_n=\frac{\chi^\alpha}{b(\mu-\imath B_\alpha)}
\Bigl(e^{-(\mu-\imath B_\alpha)nb}+\frac{nb}{L}(1-e^{-(\mu-\imath
B_\alpha)Nb})-1\Bigr)\frac{b^2 (\mu-\imath B_\alpha)^2}{2 \cosh[\mu b-\imath
bB_\alpha]-1}\ee
Substituting these solutions to the integrand (\ref{10}),
omiting the terms of the order $\mul$, summing over $n$,
rescaling the fields $\;\chi\rightarrow \sqrt{bL}\chi\;\;$
and keeping only the terms of the order ${\cal O}(b^2)$,
we get

\be I[B]\approx\int\exp\Bigl\{-\sum_\alpha
\frac{\chi^{\alpha *}\chi^\alpha}{\mu^2+B_\alpha^2}
\Bigl[ 1+\frac{b^2(B_\alpha^2-\mu^2)}{6}
+\frac{Nb^3}{24\mu}(3\mu^2-B_\alpha^2)(B_\alpha^2+\mu^2)\Bigr]
\Bigr\}D\chi^* D\chi\label{fft5}\ee

From the expression (\ref{fft5})
we see that the algorithm's error terms of the order ${\cal O}(b^2)$, which
are associated to replacement of the sum over $n$ by the integral,
explicitly depend on $N$ for fixed $b$.
Hence a more careful approach
to an analysis of systematic error of the order ${\cal O}(b^2)$ is needed.

As a rule for interesting cases (and for the model under consideration)
the following condition holds
\be \mu\ll ||B||\ee
Therefore, from the expresions (\ref{13},\ref{fft5}) one sees that
the relative systematic error for measurement of arbitrary quantity
in Slavnov's algorithm can be represented as follows

\bea &&\Delta=\Delta_1+\Delta_2+\Delta_3+\Delta_4 \nonumber\\&&
\Delta_1=Fb^2\;\; ;\;\;\Delta_2=G\frac{Nb^3}{\mu}\;\; ;\;\;
\Delta_3=H e^{-\mu L}\;\; ;\;\; \Delta_4=P\mu^2\label{pq52}\eea
where the values $F,G,H,P$ are approximately constant and depend on the
parameters $b,N,\mu$ only in the next order of perturbation theory.

Using the formula (\ref{pq52}), let us implement a more correct
investigation of the algorithm's deviation of the order ${\cal O}(e^{-\mu L})$.

In the third "experiment" we fixed $\mu=1$, $\frac{b^3 N}{\mu}=0.04$
($\Delta_2=const$, $\Delta_4=const$), and increased $N$
(decreasing the errors $\Delta_1$ and $\Delta_3$).
The results are reported in Table 3.

\vspace{1cm} \begin{center}
\begin{tabular}{|l|l|l|l|l|l|} \hline
$N$ &20&30&40&50&60\\
\hline
$b$ &0.1260 &0.1101 &0.1000 &0.0928 &0.0874 \\
\hline
$\mu L$&2.52 &3.30 &4.00 &4.62 &5.24\\
\hline
plaquette&0.5780(3)&0.5773(6)&0.5782(4)&0.5773(4)&0.5770(10) \\
\hline
\end{tabular}\label{table3}
\end{center}
{\bf Table 3:} {\small Simulation results at
$\mu=1.0$ , $\frac{b^3N}{\mu}=0.04$.}
\vspace{1cm}

From Table 3 one sees that the plaquette value
changes slowly when $N$ increases and it's alteration is not monotonic.
Such a behavior can be explained by a mutual compensation of the
errors $\Delta_1$ and $\Delta_3$.
Puting together the results from tables 1 and 3 one can conclude
that at given level of statistics  it is enough to hold $\mu L=5$ in order
to make the systematic error of the order ${\cal O}(e^{-\mu L})$ small.

In the fourth "experiment" we fixed
$\frac{b^3N}{\mu}=0.04$, $\mu L=3.3$ ($\Delta_2=const$, $\Delta_3=const$),
and decreased $\mu$ (decreasing the errors $\Delta_1$ and $\Delta_4$).
The plaquette value and the masses of $\pi$ and $\rho$ mesons were mesured.
The results of the simulations
are reported in Table 4. In Fig. 3 the plaquette values multiplied by 100
are plotted against the value of auxiliary "mass" $\mu$.

\begin{figure}
\epsfxsize=0.5\textwidth
\epsfbox{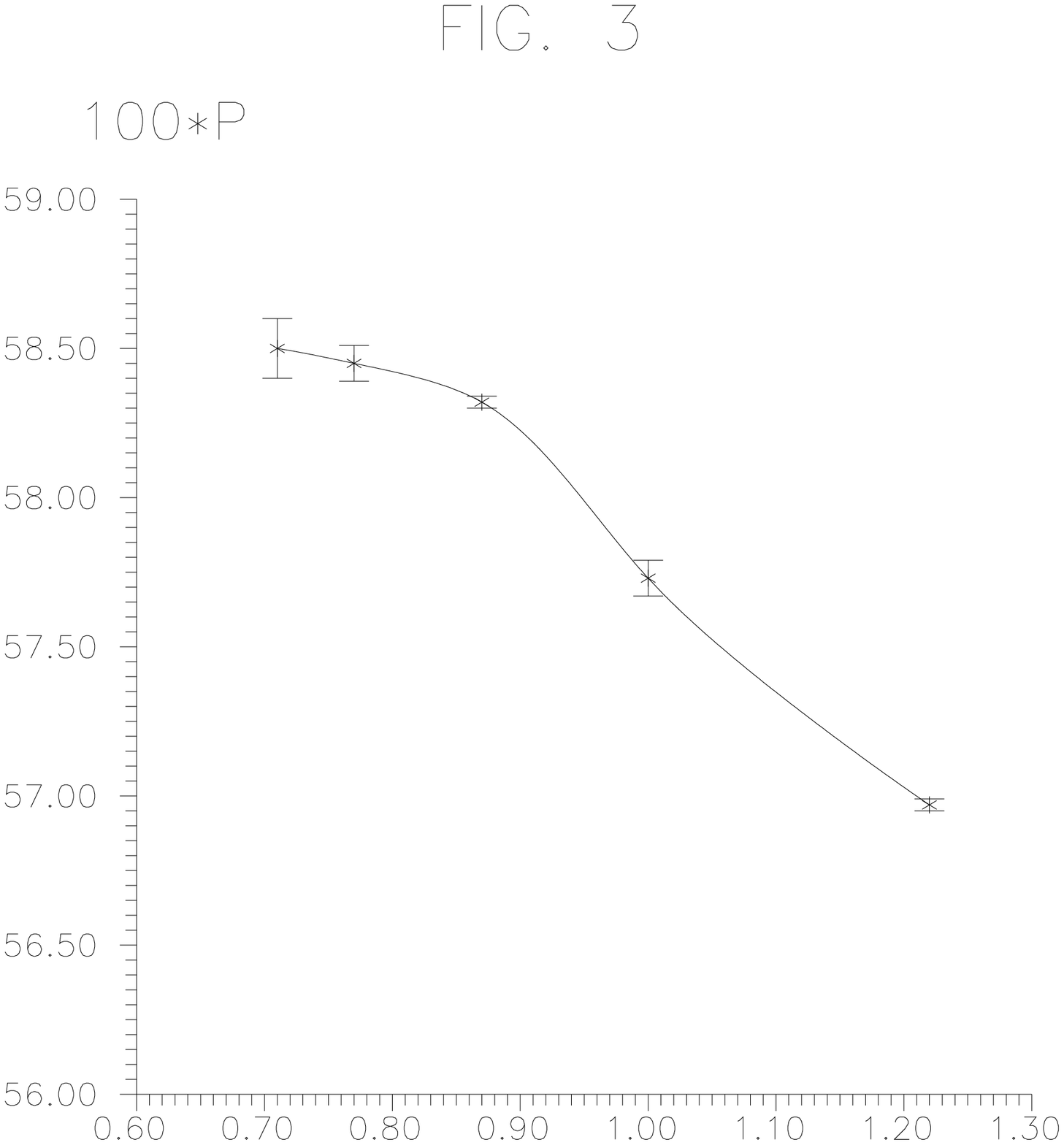}
\vspace{-1.5cm}
\Large
\hspace{7.3cm}$\mu$
\normalsize
\vspace{1.5cm}
\end{figure}

\vspace{1cm} \begin{center}
\begin{tabular}{|l|l|l|l|l|l|} \hline
$N$ & $\mu$ & $b$ &plaquette& $m_\pi$ & $m_\rho$ \\
\hline
20 & 1.22 & 0.135 & 0.5697(2)&0.93(1)&1.01(1) \\
\hline
30 & 1.0 & 0.110 & 0.5773(6) &1.02(3)&1.10(3) \\
\hline
40 & 0.87 & 0.095 & 0.5832(2)&1.10(2)&1.18(2) \\
\hline
50 & 0.77 & 0.085 & 0.5845(6)&1.16(3)&1.25(3) \\
\hline
60 & 0.71 & 0.078 & 0.5850(10)&1.15(4)&1.24(4)\\
\hline
\end{tabular}\label{table4}
\end{center}
{\bf Table 4:} {\small Simulation results for the plaquette and meson masses
at $\frac{b^3N}{\mu}=0.04$, $\mu bN=3.3$ .}
\vspace{1cm}

The simulation results for the plaquette and meson masses from Table 4
show that the measured values stabilize when $\mu$
decreases. From the data obtained we see that at this level of statistics
one may be able to do with $\mu=0.8$ to make the error of the
order ${\cal O}(\mu^2)$ small.

Let us make a conclusion to this section. We investigated the systematic
errors of Slavnov's algorithm and got the values of algorithm's parameters
at which in the tested model the errors can be considered small
at the given level of statistical precision. These values are as follows:

\be \frac{b^3N}{\mu}\le 0.03\quad ;\quad b \le 0.08\quad ;\quad
\mu bN \ge 5.0\quad ;\quad \mu\le 0.8 \label{ol1}\ee
From the conditions (\ref{ol1}) we infer that it is enough to
take $N=100$ points in auxiliary dimension to get correct results
in the tested model.
To check up this statement we made a control test run
at $N=100\; ;\; \mu=0.8\; ;\; b=0.062$.
The measured plaquette value $<P>=0.5805(11)$ within the
statistical error coincided with the result obtained from HMC in paper
\cite{LuescherE}: $<P>_{HMC}=0.5796(6)$.

\section{Conclusion.}

In this paper we performed the first investigation of systematic
error effects of Slavnov's algorithm in a realistic model.
It was shown that at this level of statistics one may be able to
do with $N=100$ points in auxiliary dimension. This is comparable
with the number of bosonic fields which was necessary
to get the correct results in the same model from L\"{u}scher's algorithm
in Ref.\cite{LuescherE}.

Note that the action (\ref{6}) and the effective bosonic action in
L\"{u}scher's formulation of the dynamical fermions problem
are similar in a sense that computational effort per update cycle
is almost the same for both algorithms if the number of bosonic fields
in L\"{u}scher's algorithm is equal to the number of points in auxiliary
dimension $N$ in action (\ref{6}) (the update of the fields
$\chi$ in the action (\ref{6}) takes less than $1\%$ of total computer time).
Accordingly both algorithms are comparable in an amount of computer time
which is necessary per full update cycle.

We observed the growth of autocorrelation times when $N$ increases.
It seems that such an autocorrelation behavior is the general feature
of multibosonic algorithms (the theoretical justification of this fact
can be found in Ref.\cite{Jeger1}).
The autocorrelation times in our simulations for $N=20$ and $40$
are almost two times less than the ones for the same numbers of bosonic fields
in L\"{u}scher's formulation measured in Ref.\cite{Jeger1},
but our MC runs were rather short and that may tend to underestimating
the algorithm's autocorrelation time. A more careful investigation of
autocorrelation behaviour in Slavnov's algorithm requires further extensive
tests on a more powerful computers.

It is interesting that the systematic errors ${\cal O}(b^2)$ and
${\cal O}(\mu^2)$ from the one side and $\mul$ from the other side
for the measured quantities compensate each other. This fact
can lead to the considerable improvement of results even for
moderate values of $N$ as it was observed in the present research.

The present performance of Slavnov's algorithm can be considerably
improved by diminishing the systematic error from ${\cal O}(b^2)$
to ${\cal O}(b^4)$. This can be put into effect by adding to the action
(\ref{6}) the terms

\be \Delta S= \sum_x\Bigl[\alpha(b,N,\mu)\chi^*(x)B^2\chi(x)+
\delta(b,N,\mu)\chi^*(x)\chi(x)\Bigr]\ee
where the coefficients $\alpha(b,N,\mu)$ and $\delta(b,N,\mu)$
can be computed theoretically (this work is under progress).
Also even-odd preconditioning can be incorporated to reduce the number
of points in auxiliary dimension $N$.

\section{Acknowledgements.}

The numerical simulations reported in this paper were done using the APE Q4
computer of von Neumann Institute of Computational Physics (J\"{u}lich, Germany).
I am very grateful to administration of von Neumann Institute and especially
to Prof. K. Schilling for this possibility.
I am very beholden to Prof. A. Slavnov (Steclov Mathematical Institute)
for continuous attention to this work and generous help.
Special thanks to Prof. K. Schilling, Dr. T. Lippert and other members of von
Neumann Institute of Computing/DESY for hospitality and numerous discussions.

This research was supported by
INTAS-RFBR under grant 95-0681,
Russian Basic Research Fund 96-01-00551,
and Presidential grant for support of
leading scientific schools 96-15-96208.

\end{document}